\documentclass[letterpaper,12pt]{article}
\usepackage{amsmath}
\usepackage{amsfonts}
\usepackage{amssymb}
\usepackage{graphicx}
\usepackage{physics}
\usepackage{latexsym}
\usepackage{epsfig}
\usepackage{pstricks}
\usepackage{stmaryrd}
\usepackage{rotating}
\usepackage[english]{babel}
\usepackage{setspace}
\usepackage[utf8]{inputenc}
\usepackage[T1]{fontenc}
\usepackage{lmodern}
\usepackage{enumitem}
\usepackage{csquotes}
\usepackage{amsthm}
\usepackage{xcolor}

\usepackage[normalem]{ulem} 
\begin{document}
\begin{center}	
\begin{LARGE}
\textbf{On Superdeterministic Rejections of Settings Independence}\\
\end{LARGE}
\end{center}

\begin{center}
\begin{large}
G. S. Ciepielewski, E. Okon and D. Sudarsky\\
\end{large}
\textit{Universidad Nacional Aut\'onoma de M\'exico, Mexico City, Mexico.}\\
\end{center}

Relying on some auxiliary assumptions, usually considered mild, Bell's theorem proves that no local theory can reproduce all the predictions of quantum mechanics. In this work, we introduce a fully local, \emph{superdeterministic} model that, by explicitly violating \emph{settings independence}---one of these auxiliary assumptions, requiring statistical independence between measurement settings and systems to be measured---is able to reproduce all the predictions of quantum mechanics. Moreover, we show that, contrary to widespread expectations, our model can break settings independence without an initial state that is too complex to handle, without visibly losing all explanatory power and without outright nullifying all of experimental science. Still, we argue that our model is unnecessarily complicated and does not offer true advantages over its non-local competitors. We conclude that, while our model does not appear to be a viable contender to their non-local counterparts, it provides the ideal framework to advance the debate over violations of statistical independence via the superdeterministic route.

\begin{flushright}
\textit{Aclar\'o que un Aleph es uno de los puntos \\ del espacio que contienen todos los puntos.}\\ ---Jorge Luis Borges
\end{flushright}

\begin{flushright}
\textit{...il s’ensuit, que cette communication va à quelque distance que ce soit. \\ Et par conséquent tout corps se ressent de tout ce qui se fait dans l’univers; \\ tellement que celui qui voit tout, pourrait lire dans chacun ce qui se fait partout...}\\ ---Gottfried Wilhelm Leibniz
\end{flushright}

\onehalfspacing
\section{Introduction}
	
John Bell proved that no local theory can reproduce all the predictions of quantum mechanics \cite{Bell1964,Bell1971,Bell1976,Bell1990}. A few years later, Alan Aspect performed the relevant experiments \cite{Aspect1981,Aspect1982} and showed the quantum predictions to be correct. Over time, more thorough experiments have been performed \cite{weihs1998, Giustina2015, Shalm, Hensen2015}, and virtually all experimental loopholes have been examined and eliminated. Since local theories do not make correct predictions for actual experiments, the most sensible conclusion is that our universe features non-local aspects. This, we hold, is the most reasonable way of interpreting Bell's result.

This position, however, has not been universally accepted. Despite both the validity and simplicity of Bell's theorem, it is not hard to find opposition and, in some cases, blatant misconceptions surrounding it. Some authors have (often erroneously) argued that their particular version of quantum theory escapes the non-locality of Bell's theorem.\footnote{In the context of Everettian Interpretations, see \cite{wallace2012}; in the context of Consistent Histories, see \cite{griffiths2011}; in that of Relational Quantum Mechanics, see \cite{RovelliEPR}.} Others have contested the validity of the non-locality conclusion by calling attention to the presence in the theorem's proof of certain implicit auxiliary assumptions. As a result, premises that in other contexts have been used and considered uncontroversial, in this context have been brought to center stage, and in some way or another challenged. The real reason behind the inconsistency, it is argued, is not locality, but one of these implicit assumptions. If that is the case, locality would not have to be rejected as a strict principle of physics, in spite of Bell's work.

The recognition that Bell's theorem makes an assumption regarding the statistical independence between settings of measuring devices and systems to be measured---an assumption we call \textit{settings independence}\footnote{This assumption has also been called measurement independence, statistical independence or the no-conspiracy, free will or free choice assumption.}---is nothing new. It was acknowledged in many places by Bell himself and discussed explicitly in \cite{Bell1977}. Nevertheless, he and others have argued that negating settings independence leads to unreasonable theories because it seems to imply a massive conspiracy that would jeopardize all of experimental science. Still, in recent years it has been argued that one could reasonably avoid Bell's theorem, and hence have a local theory which makes correct predictions, by denying settings independence \cite{Hooft2016,Hooft2017,Sanchez2018,Weinstein}.\footnote{In fact, a rather extensive and vigorous discussion has been taking place online between strong advocates of rejecting settings independence, mainly Gerard t'Hooft, and strong detractors, especially Tim Maudlin. Such an exchange has been an important motivator for this work.} However, proponents of these ideas do not seem to have successfully addressed the opposition, nor have they been able to provide a concrete model that does the trick---that is, a concrete local model, compatible with all quantum predictions. In the absence of such a model, and given the strong arguments against it, the idea of rejecting settings independence remains as far-fetched as ever.

In this work we present a concrete, fully local model (and some variations thereof) that, by violating settings independence via the superdeterministic path, is able to reproduce all the predictions of quantum mechanics. Our main objective is to provide a concrete framework where the debate over violations of statistical independence via superdeterminism could move forward. In that spirit, we describe our model in detail and try to defend it from the standard objections against superdeterministic models. We also explore additional potential problems and discuss potential lines of defense as well as possible counterarguments.

To all do this, our manuscript is organized as follows. In section \ref{bell's theorem}, we start with an overview of Bell's theorem. We pay special attention to details that will be relevant for the settings independence discussion later on. In section \ref{SI}, we analyze in depth the settings independence assumption and present the case for it. Then, we explore \emph{superdeterminism} as a path to rejecting such an assumption and discuss the worries associated with such a path. Next, in section \ref{Model}, we present the details of what we call the \emph{local pilot-wave model}, a fully local model compatible with all quantum predictions, and, in section \ref{Dis}, we explore its reasonableness and viability. We find that, while our model fares significantly better than expected regarding worries usually attributed to violations of settings independence, it does give rise to a number of unforeseen difficulties. Moreover, we point out that it performs poorly regarding theoretical virtues such as simplicity, convenience and clarity. In the end, we conclude that, while the proposed model does not succeed in providing an interesting contender to their non-local counterparts, it provides the ideal framework to advance the debate over violations of statistical independence via superdeterminism.
\section{Bell's theorem}
\label{bell's theorem}

We start this section with an overview of Bell's theorem (see \cite{Myrvold,Scholar} for thorough discussions). Then we make a few comments regarding the relation between locality and factorizability that will be relevant for the settings independence discussion later on.

\subsection{The inequality}

Consider an ensemble of pairs of particles, all created in the state that quantum mechanics describes as a singlet. Particles of each pair are sent to two spatially separated locations, 1 and 2, where spin measurements are to be performed. Let $a,b$ denote the spin directions measured in 1 and 2, respectively, and let $A,B$ stand for the corresponding results (with spin-up corresponding to $+1$ and spin-down to $-1$). Denote by $\lambda$ the complete, fundamental, state of each pair; that is, $\lambda$ contains complete information regarding all fundamental properties of the pair. If quantum mechanics is complete, $\lambda$ is the wave function; if it is not, then $\lambda$ could be the wave function supplemented with extra information or it could be something entirely different. No assumption regarding the nature of $\lambda$ is made. In particular, one does not impose for $\lambda$ to necessarily constitute ``hidden variables'' that \emph{complete} standard quantum mechanics, nor that the fundamental model is deterministic.

As one is interested in exploring the viability of local models, one assumes, with Bell, that the description under consideration is local. In particular, one assumes that the probabilities $P\left(A,B|a,b,\lambda\right)$ predicted by the model for the experiments in question are such that
\begin{equation}\label{fac}
P\left(A,B|a,b,\lambda\right) = P\left(A|a,\lambda\right) P\left(B|b,\lambda\right).
\end{equation}
This condition, usually referred to as \emph{factorizability}, was introduced by Bell to capture the idea that, for local theories, all correlations between distant systems must be locally explicable. That is, once one conditionalizes on the complete state $\lambda$, correlations between distant measurements disappear.

Next, one notes that the set of measured systems must be characterized by some (normalized) distribution over its fundamental states---call such a distribution $\rho(\lambda)$. The point is that, even though all measured pairs are described, from a quantum point of view, by the same state (i.e., a singlet), the complete description given by $\lambda$ may very well change from pair to pair. After all, we have control over the quantum state, but not over the underlying fundamental state $\lambda$. One assumes that the distribution $\rho(\lambda)$ and the settings $a$ and $b$ are statistically independent:
\begin{equation}\label{mi}
\rho(\lambda|a,b) = \rho(\lambda) ;
\end{equation}
we call this condition \emph{settings independence} (SI). Intuitively, this assumption entails that the settings $a,b$ and $\lambda$ are not correlated. In the next section, we dive deeply into the meaning of SI and the implications that come from denying it. For now, it is sufficient to say that, on the surface, the assumption seems reasonable because one can set up things in such a way that the settings of the detectors can be chosen before, during or after the particle generator has emitted the pair of particles, and they can be chosen in a myriad of extravagant and convoluted ways. Moreover, an analogous assumption is universally accepted, albeit implicitly, in all experimental scenarios across all sciences.

Given these assumptions, it can be shown that the expectation value of the product $AB$ over the whole ensemble,
\begin{equation} \label{expectation1}
E(a,b) = \int \sum_{A,B} A \, B \, \Pr(A,B|a,b,\lambda) \, \rho(\lambda|a,b) \, d\lambda,
\end{equation}
necessarily obeys
\begin{equation}
|E(a,b) + E(a,b') + E(a',b) - E(a',b')| \leq 2 .
\end{equation}
That is, all local models must satisfy this inequality. This, in a nutshell, is Bell's theorem.

It is instructive to review the proof of the theorem. Using the assumptions in Eqs. (\ref{fac}) and (\ref{mi}), one rewrites Eq. (\ref{expectation1}) as
	\begin{equation} \label{proof2}
	\begin{split}
	E(a,b) & = \int \sum_{A,B} A \, B \, \Pr(A,B|a,b,\lambda) \, \rho(\lambda|a,b) \, d\lambda\\
	&= \int \sum_{A,B} A B \, \Pr(A|a,\lambda) \, \Pr(B|b,\lambda) \, \rho(\lambda) \, d\lambda \\
	&= \int \bigg[ \sum_{A} A \Pr(A|a,\lambda) \bigg] \bigg[ \sum_{B} B \Pr(B|a,\lambda) \bigg] \rho(\lambda) \, d\lambda \\
	&= \int \bar{A}(a,\lambda) \, \bar{B}(b,\lambda) \rho(\lambda) \, d\lambda
	\end{split}
	\end{equation}
with $\bar{A}(a,\lambda)$ and $\bar{B}(b,\lambda)$ the expressions in the square brackets of the third line above. Next, one considers
	\begin{equation}\label{E-E}
	E(a,b) \pm E(a,b') = \int \bar{A}(a,\lambda) \big[ \bar{B}(b,\lambda) \pm \bar{B}(b',\lambda) \big] \, \rho(\lambda) \, d\lambda
	\end{equation}
and notes that, since $| \bar{A}(a,\lambda) | \leq 1$,
	\begin{equation}\label{first}
	|E(a,b) \pm E(a,b')| \leq \int | \bar{B}(b,\lambda) \pm \bar{B}(b',\lambda) | \, \rho(\lambda) \, d\lambda ,
	\end{equation}
and by the same reasoning, 
	\begin{equation}\label{second}
	|E(a',b) \mp E(a',b')| \leq \int | \bar{B}(b,\lambda) \mp \bar{B}(b',\lambda) | \, \rho(\lambda) \, d\lambda .
	\end{equation}
Finally, since $| \bar{B}(b,\lambda) | \leq 1$,
\begin{equation}
	|\bar{B}(b,\lambda) \pm \bar{B}(b',\lambda)| + |\bar{B}(b,\lambda) \mp \bar{B}(b',\lambda)| \leq 2,
	\end{equation}
so by adding Eqs. (\ref{first}) and (\ref{second}), and recalling that $\rho(\lambda)$ is normalized, one arrives at
\begin{equation}\label{ineq}
	|E(a,b) - E(a,b')| + |E(a',b) + E(a',b')| \leq 2.
	\end{equation}
This shows that any local theory, i.e., any theory satisfying factorizability (used in the second line of Eq. (\ref{proof2})), must also satisfy the inequality (\ref{ineq}).\footnote{This particular inequality---known as CHSH because it was derived in \cite{Clauser1969} by Clauser, Horne, Shimony, and Holt---is not the one Bell derived in his first proof. The derivation of the original inequality requires an extra assumption of exact anticorrelations for runs in which $a=b$.}

To connect this with quantum mechanics, one notes that the quantum prediction for the expectation value in Eq. (\ref{expectation1}) is given by 
\begin{equation}
E^{QM}(a,b) = -\cos(\theta)
\end{equation}
with $\theta$ the angle between $a$ and $b$. Then, if one takes $a,a',b,b'$ on the same plane, with a 90º angle between $a$ and $a'$ and $b$ and $b'$, and a 45º angle between $a$ and $b$,
\begin{equation}\label{qprediction}
|E^{QM}(a,b) - E^{QM}(a,b')| + |E^{QM}(a',b) + E^{QM}(a',b')| = 2\sqrt{2}.
\end{equation}
Since $2\sqrt{2} \nleq 2$, it is clear that the local theories under consideration make predictions for these experiments that are incompatible with those of quantum mechanics. We conclude, then, that local theories cannot always make the same predictions as quantum mechanics.

The final step is to take into account the experiments we mentioned in the introduction. Those experiments have strongly corroborated the quantum predictions, establishing, on the way, clear violations of the inequality. It seems, then, that local theories are unable to correctly describe the world we live in: our universe appears to contain non-local features.

\subsection{Factorizability and local causality}
\label{facloc}
Before exploring in detail SI, and the possibility of rejecting it, we make a few comments regarding the relation between locality and factorizability that will be relevant later on. As we saw above, the key condition used to derive the Bell inequality is factorizability. While, initially, such a condition was assumed by Bell at the onset as a defining characteristic of local theories, in later presentations of the inequalities, Bell regarded factorizability as derivable from what he called the \emph{principle of local causality} (see, e.g., \cite{Bell1990}).

According to such a principle, a model is local if the probability it assigns to $b_\chi$, the value of some \emph{beable}\footnote{The beables of a theory, a concept introduced by Bell, are those entities in it ``which are, at least tentatively, to be taken seriously, as corresponding to something real'' \cite{Bell1990}.} or property at the space-time event $\chi$, is such that
\begin{equation} \label{condprobA}
P(b_\chi|\lambda_\sigma) = P(b_\chi| \lambda_\sigma,b_\xi),
\end{equation} 
with $\lambda_\sigma$ a complete specification of the physical state on $\sigma$, a spatial slice fully covering the past light cone of $\chi$\footnote{Technically speaking, one requires $\sigma$ to be such that $\chi$ lies on its future domain of dependence, $ D^{+}( \sigma) $.}, and $b_\xi$ the value of any beable or property on an event $\xi$, space-like separated from $\chi$ and outside of the causal future of $\sigma$ (see Figure 1).
\begin{figure}[ht]
\centering
\includegraphics[height=6cm]{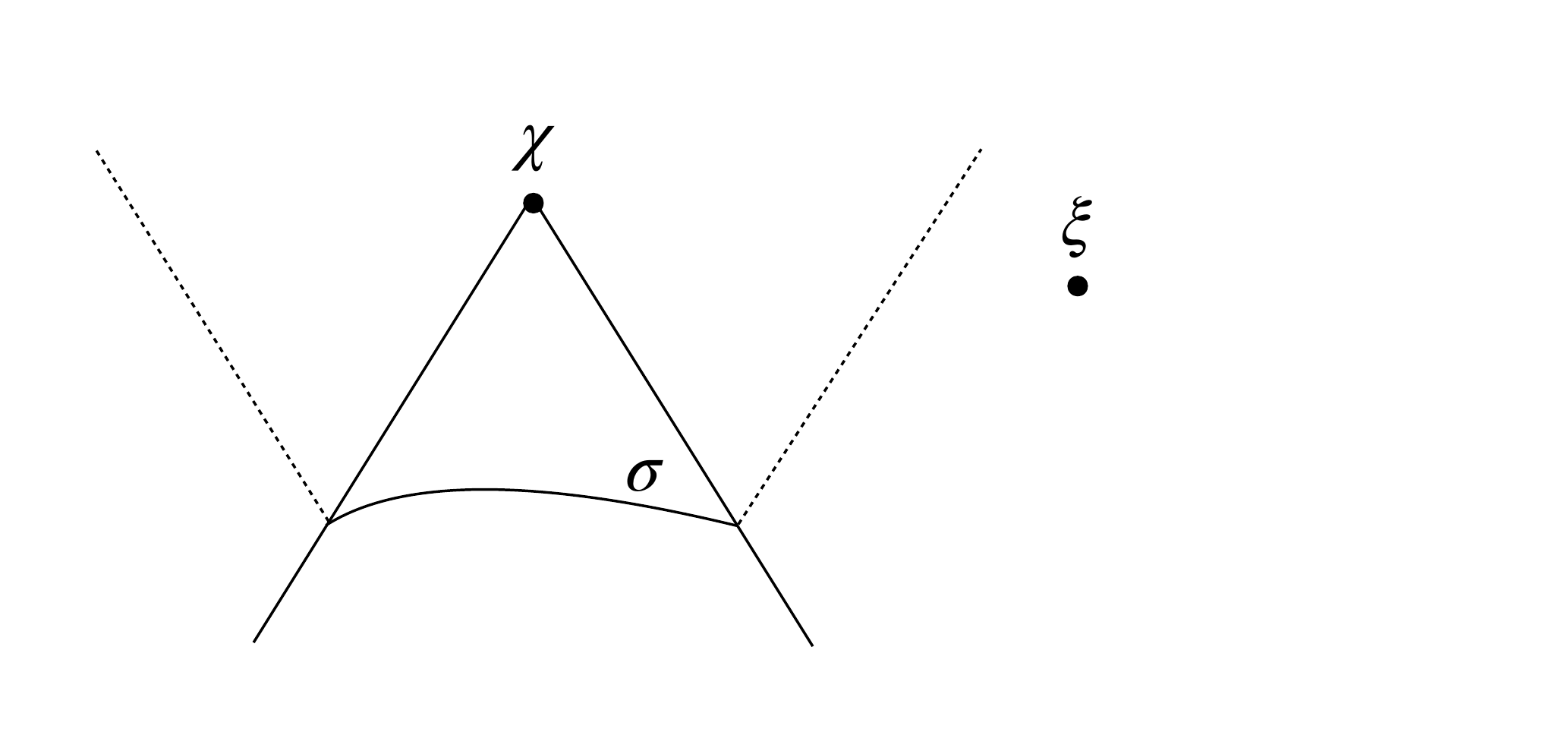} 
\caption{According to Bell's principle of local causality, a theory is local if $P(b_\chi|\lambda_\sigma) = P(b_\chi|\lambda_\sigma,b_\xi)$.}
\end{figure}
That is, for local models, if complete information on a slice of the past light cone of an event is available, then, new information regarding happenings on regions outside of the future of such a slice cannot alter the predictions of the model regarding that event. It is important to point out that this definition of locality presupposes that all beables are local (e.g., it assumes there is such a thing as $b_\chi$, a beable at $\chi$, or $\lambda_\sigma$, the complete state on region $\sigma$). In fact, it seems impossible to construct a notion of locality without assuming the existence of at least some local beables (and it is not clear what role could non-local beables play in a local model).

Bell makes two important remarks regarding this principle. First, that it is crucial for $\xi$ to lie outside of the causal future of $\sigma$. Second, that it is crucial for events in $\sigma$ to be specified completely \cite{Bell1990}. If either of these two things fails to obtain, then, even for a local theory, $\xi$ could provide information about $\chi$ (if the description in $\sigma$ is not complete, $\xi$ could complement it; if $\xi$ does not lie outside of the causal future of $\sigma$, then, in an indeterministic theory, a stochastic process in the future of $\sigma$, but in the common past of $\chi$ and $\xi$, could correlate them; for details, see \cite[p.16]{norsen2017}).

In order to explore the relation between factorizability and local causality, we consider Bell's experimental scenario (see Figure 2).
\begin{figure}[ht]
\centering
\includegraphics[height=6cm]{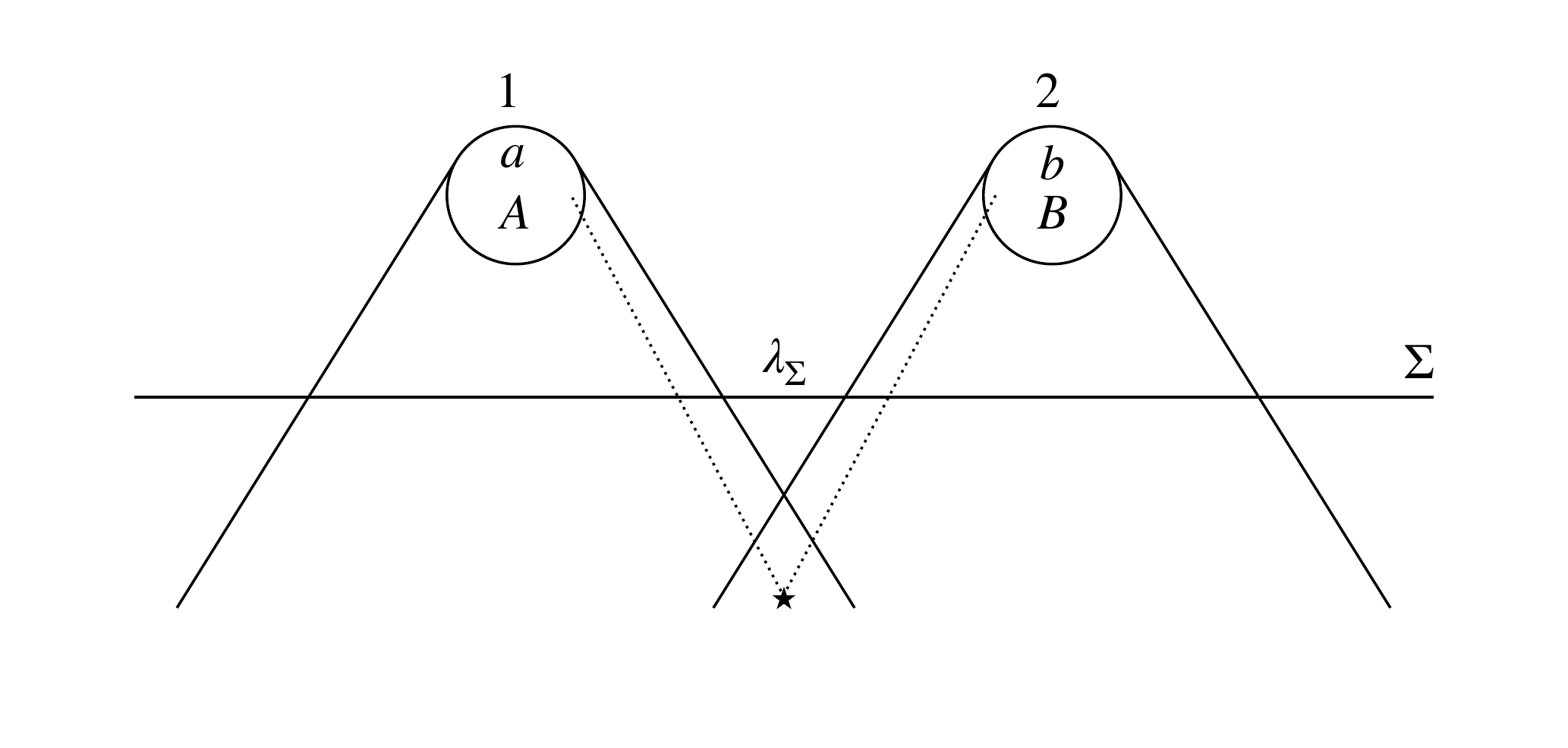} 
\caption{Space-time diagram of Bell's experimental scenario.}
\end{figure}
First we note that, while the principle considers a situation in which $\sigma$ is completely contained on and within the past light cone of $\chi$, Bell's scenario is such that complete information is provided on a full hypersurface $\Sigma$, extending all the way to infinity. This allows to consider models that contain so-called non-local beables (such as the wave function, which presumably, at least partially, describes the state, and is not associated with a particular space-time region, but to the whole hypersurface). Second, we note that, while the principle considers \emph{any} beable outside of the future of $\sigma$, in this scenario, we only consider probabilities conditionalized on the particular set of parameters, $A,B,a$ and $b$. The question, then, is if these differences could lead to trouble; that is, if the application of the modified criterion could fail in capturing local models as intended.

The answer is that there is nothing to worry about because, by allowing the introduction of information from outside of the past light cone, and by only considering conditionalizations over a few parameters, the criterion turns into a necessary, but not sufficient condition for locality. That is because, while of course all local theories are such that their probabilities are not altered by specifying additional information of the sort considered in Bell's principle\footnote{That is, information from outside of the causal future of the intersection between $\Sigma$ and the past light cone of regions 1 and 2, respectively.} (the condition is necessary), it is also the case that not all non-local theories are such that their probabilities are in fact altered by the particular new information provided by the result and setting on the other side (the condition is not sufficient).\footnote{In fact, if indeed the information on $\Sigma$ is complete, all \emph{deterministic} non-local theories will satisfy the criterion; in due time we will relax this to include all sorts of non-local theories.\label{fn}} Then, all local theories, together with some non-local ones, will satisfy it. As a result, if, with Bell, what we want is to use the principle to apply to all local theories, then the modified version will do so perfectly well. Of course, some non-local theories will be taken for the ride, but that is not a problem.

Now, to derive factorizability from local causality, we first consider the joint probability $P(A,B|a,b,\lambda_\Sigma)$, with $\lambda_\Sigma$ the complete state over $\Sigma$, and write it as follows
\begin{equation}\label{PAB}
P(A,B|a,b,\lambda_\Sigma) = P(A|a,b,B,\lambda_\Sigma) P(B|a,b,\lambda_\Sigma),
\end{equation}
Next, from the principle of local causality, we see that
\begin{equation} 
P(A|a,b,B,\lambda_\Sigma) = P(A|a,\lambda_\Sigma) \quad \mbox{and} \quad P(B|a,b,\lambda_\Sigma) = P(B|b,\lambda_\Sigma) ,
\end{equation} 
from which it follows that 
\begin{equation} 
P(A,B|a,b,\lambda_\Sigma) = P(A|a,\lambda_\Sigma) P(B|b,\lambda_\Sigma).
\end{equation}
This sure looks a lot like factorizability, but we are not there yet. The problem is that, while factorizability, as used in the derivation of the theorem, is written in terms of the state of the pair, $\lambda$, the equation above is written in terms of $\lambda_\Sigma$, which, as we said, is the state on the whole hypersurface $\Sigma$. Why is this difference important? It is because $\lambda_\Sigma$ does not satisfy the equation analogous to SI, i.e., Eq. (\ref{mi}) with $\lambda$ substituted by $\lambda_\Sigma$, that would be necessary to derive the inequality. That is, $a$ and $b$ are in general correlated with $\lambda_\Sigma$ (e.g., in a deterministic theory, $\Sigma$ determines $a$ and $b$, so they surely are not independent; and even in a non-deterministic theory, one would expect some degree of correlation).

How, then, do we obtain factorizability? To do so, we first note that $\lambda_\Sigma$ can be decomposed into four (not necessarily independent) parts: $\lambda$ (describing the physical state of the particle pair to be measured), $\lambda_a$ (influencing or determining the setting $a$), $\lambda_b$ (influencing or determining the setting $b$), and $\lambda_E$ (which includes everything else). As a result, we can write Eq. (\ref{PAB}) as follows
\begin{equation} \label{LE}
P(A,B|a,b,\lambda, \lambda_a, \lambda_b, \lambda_E) = P(A|a,\lambda, \lambda_a, \lambda_b, \lambda_E) P(B|b,\lambda, \lambda_a, \lambda_b, \lambda_E).
\end{equation}
Now, since $\lambda_a$ and $\lambda_b$ only influence the measurement through $a$ and $b$, they are redundant and can be removed. Likewise, $\lambda_E$ is irrelevant by definition, so it can also be removed. With this, we finally arrive at
\begin{equation} 
P(A,B|a,b,\lambda) = P(A|a,\lambda) P(B|b,\lambda),
\end{equation}
that is, factorizability.\footnote{Note that, by removing $\lambda_E$ from the conditional, the criterion is no longer automatically satisfied by deterministic non-local theories (see fn. \ref{fn}).}

Before moving on, it is instructive to compare our derivation of factorizability from Bell's principle of local causality to that of Bell himself in \cite{Bell1990}. As we do in Eq. (\ref{LE}), Bell divides the information in the conditional into different terms: $a$, $b$, $c$ and $\lambda$ (see Eq. (9)). According to Bell, $c$ represents \emph{values of other variables describing the experiment;} however, he assumes that $c$ and $\lambda$ give a specification which is \emph{complete}, at least for the intersection of $\Sigma$ with the union of the past light cones of regions 1 and 2 (see Figure 2). It seems, then, that $c$ contains much more than a description of the experiment and is much closer to what we call $\lambda_E$. At that point, and in contrast to what we do, Bell does not attempt to remove $c$ from the conditional and keeps it through the derivation of the theorem. Moreover, he assumes that $c$ remains constant throughout the whole experiment (see, e.g., his Eq. (11)). Given that $c$ is required to contain much more information than the experimental arrangement, the assumed constancy of $c$ seems unwarranted, and potentially problematic (we will have more to say about this point in the next section).

\section{Settings Independence}
\label{SI}

We already saw that, in order to derive a conflict between local theories and quantum mechanics, SI needs to be assumed. Initially, SI was only implicitly imposed by Bell, but a few years later Clauser and Horn realized that such an assumption was independent of locality and made it explicit (see \cite[fn. 13]{CH74}). To illustrate the fact that SI and locality are independent, Shimony, Horne and Clauser consider in \cite{SHC} a scenario in which a set of experimental data for a Bell-type experiment, displaying quantum correlations, is fabricated. The results of the set are then sent to an unscrupulous apparatus manufacturer that programs two apparatuses to display the concocted results. The settings are likewise sent to the complicit assistants of two experimentalists. Finally, when the experimentalists perform the measurements, their assistants persuade them to use the fabricated settings for each run and, when the data is recorded and analyzed, results that violate Bell inequalities are uncovered. The Bell inequalities are then violated, even though all processes involved are fully local.

As we explained above, SI consists of the assumption that the distribution of physical states that characterizes the set of measured systems, $\rho(\lambda)$, and the measurement settings, $a$ and $b$, are statistically independent: $\rho(\lambda|a,b) = \rho(\lambda)$. That is, one assumes that, if out of the whole ensemble of measured pairs, one focuses on a subensemble of runs with any particular pair of settings, then the distribution of $\lambda$ over that subensemble is the same as the distribution of the whole. It is easy to see that this is employed in the derivation of the inequality in, e.g., Eq. (\ref{E-E}).

There are a number of confusions in the literature regarding the exact nature and meaning of SI. The first we want to mention has to do with the fact that SI is often presented as a claim regarding the freedom of the experimenter to choose the settings (see, e.g., \cite{Bell1976}). This has led to the impression that SI has something to do with the existence of free will (or lack thereof). It should be clear, however, given the description given above, that SI has nothing to do with questions regarding the existence of true free will. Of course, since SI is the assumption that there is no correlation between $\lambda$ and the settings, the issue arises whether experiments can be carried out, such that SI is satisfied---and this is a very important question, to which we come back soon---but such a question is clearly independent of the free will issue.

Another possible misunderstanding might arise from the fact that $\rho(\lambda)$ is often portrayed as a \emph{probability} distribution. We believe, however, that it is helpful to keep in mind that it represents the \emph{actual} distribution of states over the measured ensemble. If $\rho(\lambda)$ is thought of as a probability distribution, then one could reasonably ask: probability conditional on what? Given that we do not have an explicit theory for $\lambda$, then it, in fact, could depend on many things, such as the temperature of the lab, the time of the experiment, etc. If so, then one could find it hard to believe, not only that it remains constant when different settings are employed, but that it remains constant over time at all. For instance, in \cite{Bell1990} Bell explicitly writes the dependence of $\lambda$ on $c$ (see his Eq. (12)). However, as we explained, and, in contrast to what Bell does in that article, we consider it unjustified to assume that $c$ remains constant throughout the experiment. If $\rho(\lambda)$ is identified, not with a probability distribution, but with the actual distribution over the measured ensemble, then the issue becomes much more tractable. Another potential problem in regarding $\rho(\lambda)$ as a probability distribution is that one could claim that, since the theory controlling $\lambda$ could be anything, then we should not impose on it the requirement that it employs classical probabilities: e.g., we should not impose on $\rho(\lambda)$ to obey classical probability rules. This, of course, becomes a non-issue once one recognizes that $\rho(\lambda)$ is basically a histogram of the actual distribution of $\lambda$ over the actual experimental ensemble. Of course, there is a relation between the details of a model, the probability it assigns to a given $\lambda$ in a particular circumstance, and the actual $\rho(\lambda)$ that is produced on a particular experiment. The point we are making is that such details are irrelevant for the construction of the inequality.

Yet another confusion regarding the nature of SI has to do with ignoring the distinction between $\lambda$, the state of a pair, and $\lambda_\Sigma$, the state over the whole hypersurface $\Sigma$. As we explained above, in general, $a$ and $b$ are correlated with $\lambda_\Sigma$, so it is clear that the analogous of SI, with $\lambda_\Sigma$ substituted for $\lambda$, is not satisfied for $\lambda_\Sigma$. The idea behind SI, then, is that, out of all the degrees of freedom over $\Sigma$, there is a subset, which we call $\lambda$, that not only completely characterizes the pair, but that can be made independent of the subsets of $\lambda_\Sigma$, $\lambda_a$ and $\lambda_b$, influencing or determining the settings. Is this an acceptable assumption? We turn to that question next.

\subsection{Justifying Settings Independence}

As we mentioned above, the key question to explore the reasonableness of SI is whether there are designs for Bell-type experiments that make it safe to assume that SI is in fact satisfied. In this regard, what one does---in physics and elsewhere in science---is to employ some sort of randomization or pseudo-randomization 
process. In clinical trials, for example, one takes special precautions to guarantee that the sample under scrutiny is representative of the whole population one is trying to study. One usually achieves this by using some sampling method that (pseudo-)randomly selects members from the target population. Likewise, in Bell-type experiments, measurement settings are usually chosen by a random or pseudo-random number generator. 
Alternatively, more complicated mechanisms have been employed for the selection of the settings. For example, these have been chosen via pseudo-random processes involving the digits of $\pi$ and popular TV shows and movies \cite{Shalm}, photons from Milky Way stars \cite{Handsteiner} or distant quasars \cite{Rauch}, and input from around 100,000 volunteers from all over the world \cite{Abellan}.

Do these elaborate methods guarantee randomness, at least at the required practical level for SI to be satisfied? It is quite reasonable to believe so. Otherwise, one would have to postulate utterly surprising correlations between seemingly independent degrees of freedom. In fact, it is well known that, unless there is a specific mechanism to maintain a given correlation, the generic dynamics of all but the simplest systems will quickly wash out correlations between any small subset of separated variables. This seems to be so even for fully deterministic theories with no fundamental randomness. Furthermore, if the pseudo-random process employed to select the settings is controlled by a given immutable sequence---say, a device that selects angles depending on the evenness or oddness of digits of $ \pi $---then the possibility of a common cause explanation for a possible correlation between the settings and $\lambda$ appears to be ruled out because nothing in the past of the system could influence the digits of $ \pi $.\footnote{This is a point raised by Tim Maudlin in an online conversation with Gerard 't Hooft.} This is what Bell himself had to say about all this in \cite{Bell1977}:
\begin{quotation}
Consider the extreme case of a `random' generator which is in fact perfectly deterministic in nature---and, for simplicity, perfectly isolated. In such a device the complete final state perfectly determines the complete initial state---nothing is forgotten. And yet for many purposes, such a device is precisely a `forgetting machine'. A particular output is the result of combining so many factors, of such a lengthy and complicated dynamical chain, that it is quite extraordinarily sensitive to minute variations of any one of many initial conditions. It is the familiar paradox of classical statistical mechanics that such exquisite sensitivity to initial conditions is practically equivalent to complete forgetfulness of them.
\end{quotation}

To illustrate the point, consider an ensemble of frictionless, perfectly elastic pool tables, each with 16 balls: one white and 15 numbered. The situation in each table is fully deterministic, with the initial positions and velocities determining the exact state of motion of all balls at all times. Assume that in all tables, all numbered balls are placed at rest at identical positions and that all white balls start with the same speed, but different directions. The ensemble of tables, then, is described by a distribution over such initial directions. Now, it is clear that there are strong correlations between all tables in the ensemble, as the complete situation is fully parametrized by time and the initial angle. In particular, there is a simple, perfect correlation between all tables that is easy to write: the sum of the kinetic energies of all balls in each table will be the same for all tables, at all times. Imagine, however, that we exclude from consideration a couple of balls, say, those with numbers 14 and 15. What kind of correlation will we find among the states of the different tables, e.g., regarding the sum of kinetic energies of the remaining balls? It seems that the perfect correlation we had before will be degraded dramatically. What happens if we now remove even more balls from consideration, say those in the interval 8-15. The degradation in the correlations would seem to increase and, probably, it would be hard to find any simple correlation of any kind. What happens, finally, if we only consider a couple of balls, say, those with numbers 1 and 2. Generically, one would expect for any correlations left to be impossible to detect in practice. The idea is that there is a huge amount of degrees of freedom, other than balls number 1 and 2, that would provide sufficient ``padding'' between them, so there is no reason to expect a correlation.

Going back to the Bell case, think of the ensemble of pool tables as analogous to the ensemble of experimental runs and think of the kinetic energies of balls 1 and 2, as analogous to $\lambda$ and the settings. Then, in the same way that one does not expect any correlations between the kinetic energies of balls 1 and 2, in the Bell case, which clearly involves something much more radical than the pool tables, with an inconceivably large number of degrees of freedom being ignored, one does not expect any correlations between the state $\lambda$ and the settings (or whatever is used to determine them, e.g., photons from Milky Way stars or input from people from all over the world). And this is so even if the universe is fully deterministic, with correlations originating since the beginning---just as in the pool tables example above. It seems quite reasonable, then, to assume that the settings are sufficiently independent of $\lambda$, making SI accurate enough for the inequality to follow. That is, $\lambda_E$ is expected to be so big, that it provides sufficient ``padding'' between $\lambda_a$ and $\lambda_b$ (on the one hand) and $\lambda$ (on the other) so there is no reason to expect correlations between $a$ and $b$.

All these arguments in favor of SI seem fairly solid; still, some have tried to resist them. In what follows, we consider one proposed avenue for rejecting SI: superdeterminism. We then delineate some of the worries usually associated with such a path.

\subsection{Superdeterminism}
\label{RSI}

The arguments in favor of SI appear to be quite strong. Still, strictly speaking, they are not conclusive. After all, it might be that the ideas about effective randomness described above, which appear reasonable, are ultimately wrong. Or that, for some unknown reason, they do not apply to the particular Bell-type cases under consideration. That is, it might be that, no matter how hard we try to make $\lambda$ and the settings independent, they are still correlated. A conspiracy appears to be required for this to happen, but the truth is that there is no way to prove that no conspiracy of that sort is behind the breaking of the Bell inequality. That is, a non-local world and a local-but-conspiratorial one can be completely indistinguishable at the empirical level. Moreover, a conspiracy might appear to be required only because we have not been able to uncover the mechanism that preserves the required correlations. After all, $\lambda$ and the settings do share a common past (and future) that could explain a breakdown of SI. What about the argument that no common cause could explain the correlations once we select settings using a predetermined, timeless sequence, such as $\pi$? Such an argument is not conclusive because, in the design of any experiment using randomizers, there will be something or someone arranging things up in a way that correlates the settings to that particular sequence; hence, it is still possible to deny SI because there could be a correlation between $\lambda$ and the {\it decision} to arrange things that way.

This sort of considerations, often enhanced by strong intuitions regarding the necessity of preserving locality, has led to efforts to reject SI. Broadly speaking, two strategies have been proposed to accomplish this. The first one, usually referred to as \emph{superdeterminism}, consists of denying SI by postulating that something in the (possibly distant) past ensures that the settings and $\lambda$ are correlated. The other option is to consider the possibility of \emph{retrocausality}, i.e., that causal influences can travel to the past. If so, one could use these influences to the past to explain a breakdown of SI. Retrocausality was first considered in \cite{Cos1953} in the context of the EPR argument and years later presented as an alternative to the non-local consequence of Bell's theorem \cite{Cos1977}. If causal influences are allowed to travel to the past, then the measurement of, say, the spin of particle 1 along $a$, might influence the state of $\lambda$ when it was created. Then, such an influence on $ \lambda $ would have an impact on particle 2. By this causal zigzag, the Bell statistics would be locality explained (see \cite{Whar} for a recent review). In this work, we do not explore the retrocausal alternative at all; in fact, we assume that no retrocausal mechanisms are at play and focus exclusively on superdeterminism.

As we explained above, superdeterminism maintains that $ \lambda $ and the settings are correlated because of certain conditions of the past. And this is postulated to be so, even if the method for choosing the settings is so complicated, and involves such a high sensitivity to small perturbations, that ordinarily it would be rendered as effectively random. That is, regardless of the randomization method employed, superdeterminism postulates the settings to be determined by the conditions at some past time, and to be distributed in such a way that they always produce the correct statistics in Bell-type experiments. For instance, in the case in which the settings are chosen employing photons from quasars, such photons would have to be perfectly coordinated with the states of the measured pairs to guarantee, via purely local processes, that the correct results are always produced.

In recent years, this position has gained some notoriety because of Gerard 't Hooft’s endorsement within his cellular automaton program \cite{Hooft2016}. For years, he has been trying to come up with a modification of quantum theory that is deterministic and fully local---among other characteristics (see \cite{Hooft2017} for a list of his theoretical requirements). The idea is to maintain locality by denying SI via superdeterminism \cite[p. 14]{Hooft2017}. What reasons does he have to deny SI? He has said many things over the years. For example, that determinism is incompatible with SI \cite[p. 2-3]{Hooft2011}, that SI assumes the existence of free will \cite[p. 3-4]{Hooft2013}, that the postulation of a new law of nature is enough to violate SI \cite[p. 12]{Hooft2013}, or that the evolution laws of the system might be involved in the violation \cite[p. 10]{Hooft2009}. As it should be clear by now, the first two points are red herrings that have nothing to do with SI (see section \ref{SI} above), while the last two are more like aspirations or strategies that he would like to see fulfilled than arguments in favor of his position.

There are, however, strong intuitions against superdeterminism. For instance, this is what Shimony, Horne, and Clauser had to say about the approach \cite{SHC}:
\begin{quotation}
We cannot deny such a possibility. But we feel that it is wrong on methodological grounds to worry seriously about it if no specific causal linkage is proposed. In any scientific experiment in which two or more variables are supposed to be randomly selected, one can always conjecture that some factor in the overlap of the backward light cones has controlled the presumably random choices. But, we maintain, skepticism of this sort will essentially dismiss all results of scientific experimentation. Unless we proceed under the assumption that hidden conspiracies of this sort do not occur, we have abandoned in advance the whole enterprise of discovering the laws of nature by experimentation.
\end{quotation}

It seems, then, that there are important hurdles for the superdeterminism program. To begin with, it is true that the arguments in favor of SI are not decisive. However, the mere possibility of these arguments being wrong does not automatically make the skeptic position interesting. In the absence of an explanation of why one could expect the arguments to fail, or a concrete physical model in which they do so, the skeptic position seems weak. It is also true that a non-local world, and a local-but-conspiratorial one, can be empirically indistinguishable, but that, by itself, is not impressive; all sorts of extravagant worlds that are indistinguishable from ours can easily be concocted (e.g., one in which inert pink dragons appear in the sky when nobody is looking). In sum, in the absence of a workable superdeterministic model, the approach seems to lose all strength.

Moreover, there are strong intuitions to the effect that, even if superdeterministic models could be built, they would require such complex initial conditions that they would be useless in practice. In other words, the initial state required to explain the correlations would need such a level of detail in its prescription---involving the specification of such a large number of quantities to such levels of precision---that, even if possible in principle, these models would be unworkable in practice (see, for instance, \cite{Kronz}). 

Another worry associated with the construction of a superdeterministic model has to do with a potential lack of explanatory power.\footnote{Regarding the confluence of fine-tuning, Bell correlations, and explanatory power, the influential \cite{WS} shows that every causal model that can reproduce no-signaling, Bell-inequality-violating correlations, must violate a condition called \emph{faithfulness}, which can be read as a condition of no fine-tuning. It is important to point out, however, that faithfulness is in fact considerably stronger than what critics of superdeterminism have in mind when they talk about fine-tuning. To see this, we note that, while faithfulness judges as fine-tuned frameworks such as standard quantum mechanics or pilot-wave theory, the worry regarding superdeterminism is precisely that it could introduce a sort of fine-tuning not present in, e.g., standard quantum mechanics or pilot-wave theory.}  In particular, it is feared that such models would not be able to provide real explanations because all explanations would boil down to properties of the initial state. That is, superdeterministic models would only be able to assert that ``things are as they are because the initial state was what it was,'' period. Moreover, it seems that all initial states required to provide explanations would have to be fine-tuned, i.e., they would have to be chosen within a set of measure zero (given a reasonable measure over the space of initial states). As a result, the explanations offered would be that ``things are as they are because the initial state was what it was (even though such an initial state was extremely unlikely).'' That does not seem like a satisfactory explanation. What is more, one would need a separate, completely independent explanation of this sort for each different Bell-type experiment. To say the least, that looks like an extremely inefficient way to try to accommodate all the empirical evidence in favor of violations of the Bell inequality.

Finally, it seems that the proposed rejection of SI in order to save locality would, in one stroke, jeopardize all experimental science. That is because an assumption analogous to SI is universally posited in all experimental scenarios across all sciences, so a violation of SI would imply that the analogous assumptions would also need to be questioned. If so, the conclusions of all sorts of experiments would have to be invalidated (e.g., all results from randomized clinical trials would have to be invalidated).\footnote{This objection is what, presumably, Shimony, Horne, and Clauser had in mind in the passage quoted above when they claim that ``skepticism of this sort will essentially dismiss all results of scientific experimentation.''} If, on the other hand, the idea is that SI is only violated for certain experiments, such as Bell-type ones, then, in the absence of a detailed explanation of why this is so, the objection becomes plain silly---any unwanted experimental outcome could be discarded this way.

In sum, the fact that the arguments in favor of SI are not conclusive, together with strong intuitions regarding the necessity to preserve locality, have fueled efforts to reject SI. One of such efforts involves so-called superdeterminism: the idea that correlations can always be locally explained by considering conditions of the past. There are, on the other hand, strong intuitions against the viability of a model of this kind. First, it has been argued that, in the absence of concrete superdeterministic models, the skepticism behind the project loses power. Second, that even if such models could be constructed, the specification of the required initial states would be so complex that the model would be useless in practice. Third, it is feared that such models would lack all explanatory power. Fourth, and lastly, superdeterminism appears to jeopardize all experimental science. We believe that the discussion between these two sets of intuitions has been hindered, in part, by the lack of specific models that could serve as frameworks where ideas can be examined in detail. However, as far as we know, up to now it was unclear whether a manageable SI violating model can in fact be built.

In the next section, we present a model (and some variations thereof) that achieves in detail all of the wishes of those seeking to trade SI for strict locality. That is, we present a model that is fully local, involves violations of SI via the superdeterministic path and reproduces all the predictions of quantum mechanics---including violations of the Bell inequalities. Moreover, we claim that our model fares significantly better than expected regarding the worries usually attached to superdeterministic models described above. In particular, we show that: i) the model can, in fact, be constructed; ii) the conditions on the required initial state are not only not unmanageable, but rather surprisingly simple; iii) as a result of such simplicity, the model has resources to retain explanatory power and iv) because it is able to explain why SI is violated only for Bell-type experimental scenarios, and not generically, the model does not threaten all of experimental science.

Still, we find that the model gives rise to a number of unexpected complications and we find that it performs rather poorly regarding theoretical virtues such as simplicity, convenience and clarity. As a result, we end up concluding that our model---and its variations---do not constitute strong contenders to existing non-local frameworks. Moreover, we provide some arguments to the effect that, any model that achieves what ours does, will have to share many of its central features. If this is true, our model offers a quite general ground where the previous alluded discussion might happen, and out of which rather generic conclusions might be extracted.

\section{The local pilot-wave model}
\label{Model}
In this section, we present what we call the \emph{local pilot-wave model} (LPW), which is a fully local model that is capable of reproducing all quantum predictions. In order to describe it, it is useful to first consider standard, non-relativistic de-Broglie-Bohm pilot-wave theory. In such a framework, the complete characterization of an $N$-particle system is given by its wave function $\Psi (y, t)$, defined over the configuration space $Q$ with coordinates $y$, together with the actual positions of the particles $\{\mathbf{X}_1 (t),\mathbf{X}_2 (t),\dots,\mathbf{X}_N (t)\}$. The positions of the $N$ particles can be represented by a single point, $Y (t) \in Q$, the position of the so-called universal pilot-wave particle.

The wave function is postulated to satisfy at all times the usual Schr\"odinger equation,
 \begin{equation}
 i \hbar \frac{\partial \Psi (y, t)}{ \partial t } = - \sum_{k=1}^{3N} \frac{\hbar^2}{2 m_k} \frac{\partial^2 \Psi (y, t)}{\partial y_k^2} + V(y)\Psi (y, t) ,
 \end{equation}
and the positions to evolve according to the so-called guiding equation, 
\begin{equation}
\frac{d Y_k(t)}{dt} = \frac{\hbar}{m_k} \left. \text{Im}\left[ \frac{\frac{\partial \Psi (y, t)}{\partial y_k }}{ \Psi (y, t)} \right] \right\rvert_{y=Y(t)}.
\label{guide}
\end{equation}
The theory is clearly non-local since, according to the guiding equation, the velocity of a given particle at a given time may depend on the position of all the other particles at that same time. Note, however, that this non-locality only arises if the wave function is entangled. That is, the position of a particle influences the velocity of another, only if the wave function of those two particles is entangled. Otherwise, the non-local behavior does not arise. In a Bell experiment scenario, the non-locality due to the entanglement of the pair of particles explains the observed correlations. In particular, assuming particle 1 is measured first, the setting on such a measurement, by affecting particle 1, also non-locally affects the behavior of particle 2, leading to violations of the Bell inequality.

The specification of the pilot-wave theory is finalized with the so-called statistical postulate, which imposes that, at some initial time $t_0$, the wave function was arbitrarily chosen to be $\Psi_0$ and the positions of the particles were \emph{randomly} selected according to a probability distribution given by $|\Psi_0|^2$. This implies, as a result of the evolution laws, that particles at any other time $t$ are distributed according to $|\Psi_t|^2$. Such a state is known as a state of quantum equilibrium. Furthermore, as a result of the equations of motion and the statistical postulate, the theory imposes strict, in principle, restrictions on the acquisition of information regarding the position of the particles beyond quantum equilibrium. As a result, this theory is empirically equivalent to standard quantum mechanics.
 
Since both of the evolution equations described above are deterministic, the pilot-wave framework can be said to be deterministic. However, as we just saw, the theory is not totally free of randomness. The statistical postulate implies an initial random event: the selection of the positions of the particles. In a sense, then, all the randomness displayed by the theory at the empirical level is pushed back to a single, master chancy event at the initial time $t_0$. After that, everything evolves deterministically. It is worth mentioning that in \cite{Valentini} Antony Valentini argues for the condition of quantum equilibrium to arise, not as a result of the statistical postulate, but as a result of some sort of dynamical relaxation over time. Serious doubts, however, have been expressed regarding the viability of that idea, and even about the possibility of constructing a modified theory in which such an idea might be realized \cite{ward}.
 
Before moving on, we comment on the ontology of pilot-wave theory. It consists of a non-local beable, represented by the wave function, and a set of local beables, the point particles. Out of those, the following mass density over physical space (with coordinates $\mathbf{x}=(x_1,x_2,x_3)$) can be constructed:
\begin{equation}
\rho_\mathbf{x}(t) = \sum_{i=1}^N m_i \delta (\mathbf{x}-\mathbf{X}_i(t)).
 \end{equation}
 
Now, we would like to show how a ``trick'' can be employed to transform standard pilot-wave theory into a \emph{fully local} framework. Moreover, we will use such a framework, and its variations, to exemplify different ways in which, by violating SI, a local theory can lead to violations of Bell inequalities. We start with the first (and central) version of the model, which we call $\text{LPW}_f$ (with $f$ standing for \emph{field} version). The framework we have in mind contains, at each point $\mathbf{x}$ in physical space, a set of internal degrees of freedom. Such internal degrees of freedom are represented by a wave function over an \emph{internal} $3N$-dimensional ``configuration space'' $C$ with coordinates $z_{\mathbf{x}}=(z^1_\mathbf{x},z^2_\mathbf{x}\ldots)$, and a point $Z_{\mathbf{x}}(t) \in C$. That is, this theory contains at each point in physical space a complete $N$-particle pilot-wave system. The state of the system is then fully represented by a wave function \emph{field}, $\Phi_\mathbf{x}(z_\mathbf{x},t)$, and a position \emph{field} $Z_\mathbf{x}(t)$ (which might also be thought of as $N$ position fields on an internal 3-dimensional space $\{\mathbf{W}_\mathbf{x}^1 (t),\mathbf{W}_\mathbf{x}^2 (t),\dots,\mathbf{W}_\mathbf{x}^N (t)\}$).

As for the dynamics, this framework contains, as standard pilot-wave theory, two evolution equations: a Schrödinger equation for the wave function field,
 \begin{equation}\label{M1}
 i \hbar \frac{\partial \Phi_\mathbf{x} (z_{\mathbf{x}},t)}{ \partial t } = - \sum_{k=1}^{3N} \frac{\hbar^2}{2 m_k} \frac{\partial^2 \Phi_\mathbf{x} (z_{\mathbf{x}},t)}{\partial (z^k_\mathbf{x})^2} + V(z_\mathbf{x})\Phi_\mathbf{x} (z_{\mathbf{x}},t) ,
 \end{equation}
and a guiding equation for the position field, 
\begin{equation}\label{M2}
\frac{d Z^k_\mathbf{x}(t)}{dt} = \frac{\hbar}{m_k} \left. \text{Im}\left[ \frac{\frac{\partial \Phi_\mathbf{x} (z_{\mathbf{x}},t)}{\partial z^k_\mathbf{x} }}{ \Phi_\mathbf{x} (z_{\mathbf{x}},t)} \right] \right\rvert_{z_\mathbf{x}=Z_\mathbf{x}(t)}.
\end{equation}
That is, the internal degrees of freedom at each point $\mathbf{x}$ behave in every way as an $N$-particle pilot-wave system. The framework as a whole, on the other hand, is very different from standard pilot-wave theory. In particular, the theory only contains local beables, represented by $\Phi_\mathbf{x}(z_{\mathbf{x}},t)$ and $Z_\mathbf{x}(t)$, and the evolution is ultra-local, i.e., all the information needed to evolve the degrees of freedom at a point $\mathbf{x}$ is fully contained in it.

A final element needed to fully construct the theory is a specification of a mass density in physical space given by
\begin{equation}\label{MD}
\rho_\mathbf{x}(t) = \sum_{i=1}^N m_i \delta \left(\mathbf{x}-\mathbf{W}_\mathbf{x}^i(t)\right).
 \end{equation}
That is, of all the internal degrees of freedom at a point $\mathbf{x}$, only very few translate into the presence of mass at such a point. In particular, a point in physical space is populated only if any of the point particles in its internal 3-dimensional space is located at a point with those same coordinates.

Before moving on, we note that it is possible to generate an equivalent ``particle'' version of the model, we call it $\text{LPW}_p$, in which one attaches an ``internal pilot-wave system'' to each of the $N$-particles that make up the system (rather than to all points in space). Each particle then moves in physical space following its own image contained in its internal system. We will see that $\text{LPW}_p$ has both advantages and disadvantages with respect to $\text{LPW}_f$. We take the latter as central in order to make the connection with 't Hooft cellular automaton more direct (see next section).

Going back to $\text{LPW}_f$, we see that, for generic initial conditions $\Phi_\mathbf{x}(z_{\mathbf{x}},t_0)$ and $Z_\mathbf{x}(t_0)$, the model will surely not violate SI nor, being local, Bell inequalities. It is easy to see, however, that a very special set of initial conditions, a set that seems to be of measure zero for any reasonable measure over the space of initial conditions, will exactly reproduce the behavior of standard, non-local, pilot-wave theory. If the initial conditions happen to be \emph{homogeneous}, that is, if they are such that both $\Phi_{\mathbf{x}}$ and $Z_{\mathbf{x}}$ are exactly the same at every point $ \mathbf{x} $ in physical space, then, even though the evolution is completely local, the behavior of the system exactly mimics that of standard pilot-wave---including the capacity to violate Bell inequalities.\footnote{$\text{LPW}_p$ would show this behavior if every particle has the same internal wave function and configuration.} This is because the independent internal $N$-particle pilot-wave systems at different points will ``magically'' coordinate and display all sorts of non-local correlations. It is as if every point in space could be thought of as a supercomputer that is running a perfectly detailed simulation of the entire universe. As a result, every point in space ``knows'' everything there is to know about the whole history of the universe.\footnote{We find it amusing that our model shares some features with Leibniz's Monadology. For instance, when homogeneous initial conditions are present, as a monad, each spatial point is a ``perpetual living mirror of the universe.'' Also, as Leibniz's pre-established harmony, homogenization in our model makes causality redundant. There are differences, of course. For one thing, our model is about physical entities, not psychic ones.}

All this is a concrete example of how the selection of fine-tuned initial conditions can be employed to violate both SI and Bell inequalities. Note, however, that the specification of initial conditions required to achieve a breakdown of Bell inequalities is not as complicated as one might have thought. Sure, it involves an enormous amount of non-local correlations, but it does not involve an unmanageable amount of complexity (more on this in section \ref{Dis}). To see that SI is indeed violated once homogeneous conditions are provided, we note that the state $\lambda$ of the particles involves the specification of both $\Phi$ and $Z$ in the region where the particles are created. But the homogeneity of initial conditions implies that such $\Phi$ and $Z$ determine such fields everywhere, including in the regions where $a$ and $b$ are located. We conclude that, in this case, $\lambda$, $a$ and $b$ cannot be independent.

It is interesting to note that this mechanism to break the Bell inequalities is in a certain way analogous to the way in which, in pilot-wave theory, all indeterminism is pushed back, via the statistical postulate, to the initial conditions, after which the evolution is fully deterministic. In the present case, all the non-locality is inserted into the initial condition; after that, the evolution is fully local.\footnote{The extent of the analogy is limited to the fact that a certain non-intuitive feature of the theory is moved from the time evolution to the initial conditions, but the particular feature involved in each case is of course quite different.} It is also worth mentioning that a trick analogous to the one employed to construct our model can be used to convert any deterministic non-local theory, even one with superluminal signaling, into a local one. If the local theory is then provided with homogeneous initial conditions, then it will exactly mimic the behavior of the non-local theory, including, if it is the case, its capacity to violate Bell inequalities.

As we saw above, $\text{LPW}_f$ is such that, if homogeneous initial conditions are selected, then, by violating SI, the model is able to break the Bell inequalities. We also note that the set of initial conditions leading to such behavior is extremely easy to characterize: the initial conditions must be homogeneous. This opens up the possibility of introducing such a succinct characterization as a constraint on the allowable initial conditions (see \cite{Weinstein} for similar ideas). That is, if we stipulate, as an extra postulate of the model, that the initial conditions must satisfy
\begin{equation}\label{Hom1}
\nabla_\mathbf{x} \Phi_\mathbf{x} = 0 
\end{equation}
and
\begin{equation}\label{Hom2}
\nabla_\mathbf{x} Z_\mathbf{x} = 0
\end{equation}
(with $\nabla_\mathbf{x} = \left( \frac{\partial}{\partial x_1},\frac{\partial}{\partial x_2},\frac{\partial}{\partial x_3}\right)$), then, for \emph{generic} initial conditions, the model would reproduce the behavior of standard pilot-wave theory and the Bell inequalities would be violated. We call this version of the model with constraints $\text{C-LPW}_f$.\footnote{We note that for $\text{LPW}_p$, the corresponding constraints would be less elegant. Instead of a simple pair of equations, such constraints would require, for the full set of particles $\lbrace 1, 2, ....... n...., N\rbrace $, that the internal wave functions and configuration are independent of $n$.} Following the analogy between randomness and non-locality mentioned above, these constraints, which introduce (effective) non-locality into an otherwise fully local theory, are analogous to the statistical postulate, which introduces indeterminism into an otherwise fully deterministic theory. 
 
For the final variation we explore the possibility of generating non-local correlations \emph{dynamically}. The idea is similar to what Valentini has in mind when he tries to dispense with the statistical postulate and show that quantum equilibrium can be achieved dynamically. What we propose is to introduce changes into the dynamics in such a way that any gradients present in the initial conditions are dynamically suppressed.\footnote{Note that this breaks the analogy with Valentini's ideas because he considers quantum equilibrium as arising dynamically, but without any change in the dynamical laws.} This could be achieved
by the following modified evolution equations
 \begin{equation}\label{lap1}
 i \hbar \frac{\partial \Phi_\mathbf{x} (z_{\mathbf{x}},t)}{ \partial t } = - \sum_{k=1}^{3N} \frac{\hbar^2}{2 m_k} \frac{\partial^2 \Phi_\mathbf{x} (z_{\mathbf{x}},t)}{\partial (z^k_\mathbf{x})^2} + V(z_\mathbf{x})\Phi_\mathbf{x} (z_{\mathbf{x}},t) + i \kappa \nabla_\mathbf{x}^2 \Phi_\mathbf{x} (z_{\mathbf{x}},t) ,
 \end{equation}
\begin{equation}\label{lap2}
\frac{d Z^k_\mathbf{x}(t)}{dt} = \frac{\hbar}{m_k} \left. \text{Im}\left[ \frac{\frac{\partial \Phi_\mathbf{x} (z_{\mathbf{x}},t)}{\partial z^k_\mathbf{x} }}{ \Phi_\mathbf{x} (z_{\mathbf{x}},t)} \right] \right\rvert_{z_\mathbf{x}=Z_\mathbf{x}(t)} + \kappa \nabla_\mathbf{x}^2 Z_\mathbf{x}(t) ,
\end{equation}
with $\nabla_\mathbf{x}^2 = \frac{\partial^2}{\partial x_1^2}+\frac{\partial^2}{\partial x_2^2}+\frac{\partial^2}{\partial x_3^2}$ 
and $\kappa$ a real parameter (we call this model with dynamical relaxation $\text{R-LPW}_f$). The idea, then, is to add terms to Eqs. (\ref{M1}) and (\ref{M2}) that produce some sort of heat equation that could lead to homogenization of $\Phi_\mathbf{x}$ and $Z_\mathbf{x}$ over physical space.\footnote{We do not see a simple way in which a similar trick could be implemented in the context of $\text{LPW}_p$.}

A few comments are in order. First, in order to guarantee that $\nabla_\mathbf{x} \Phi_\mathbf{x} $ and $\nabla_\mathbf{x} Z_\mathbf{x} $ are driven to zero, and not simply to a constant, we assume that the space over which the model is defined is closed and without boundaries (i.e., like a 3-torus). It is well known that, under such conditions, the heat equation drives any initial temperature distribution towards a fully homogeneous state. Second, the Laplacian in these equations must compete with the rest of the terms, so it is not clear that homogenization will be achieved. However, it is reasonable to assume that if $\kappa$ is large enough, then the other terms can be ignored until approximate homogenization is accomplished. After that, one can expect that the Laplacian terms could be ignored and that the dynamics approximate that of Eqs. (\ref{M1}) and (\ref{M2}).\footnote{We must admit that it is not clear how the system behaves when the various terms in the RHS of Eqs. (\ref{lap1}) and (\ref{lap2}) become of similar magnitude.} Third, the diffusion terms added above, just as those in the heat equation, permit instantaneous propagation, so they are non-relativistic. The existence of an acceptable relativistic heat equation is an interesting mathematical question on which, as far as we know, no definite consensus has been reached. However, we must note that there is not only an active program searching for such relativistic version (e.g., \cite{H1,H2,H3,H4,H5}), but also that, to the extent that nature is susceptible to a mathematical description, it is clear that something like that should exist. That is because both special relativity and the effect of heat diffusion are, without doubt, a part of nature. Thus, a version of the above mathematical trick, involving limited speed of propagation, must exist.

Finally, we note that, even if it is the case that, out of arbitrary initial conditions, these equations lead to a suppression of $\nabla_\mathbf{x} \Phi_\mathbf{x} $ and $\nabla_\mathbf{x} Z_\mathbf{x} $, the system will not exactly mimic a pilot-wave one. There are at least two reasons for that. The first is that, even if the statistical postulate is enforced, the dynamics while the gradients are not zero will not maintain quantum equilibrium. As a result, the dynamics after the gradients are suppressed will mimic a pilot-wave system in which quantum equilibrium is not attained, so the model will not reproduce the predictions of standard quantum mechanics (unless, of course, Valentini is correct and equilibrium is attained dynamically). The other reason is that in all these diffusion equations, like in the heat equation, the approach to equilibrium, which in this case is what generates homogeneity, is asymptotic. As a result, at any finite time, some level of inhomogeneity will remain in the system and the model will not really be empirically identical to a standard pilot-wave model. This, of course, means that this model would be able to produce novel predictions, which, at least in principle, could be experimentally explored. For instance, if the experiments are repeated a sufficiently large number of times, one might expect some failures in the exact correlations in Bell experiments when the spin of the two particles making up the singlet are measured along the same axis or in the perfect correlations in the three-particle GHZ set-up \cite{GHZ}.\footnote{One should take these perfect correlations as a sign of the robustness needed to really achieve full agreement with the predictions of quantum mechanics, as achieved by $\text{C-LPW}_f$)}. At any rate, it is important to keep in mind that if this model can be built, even if it does not reproduce the predictions of standard quantum mechanics, it would be able to break the Bell inequalities by purely local means, out of completely arbitrary initial conditions and without imposing any constraints.

To sum up, we have constructed a fully local model that, when supplemented with homogeneous initial conditions, exactly reproduces all the predictions of standard pilot-wave theory---including its capacity to violate the Bell inequalities. This is possible because the constructed model does not satisfy SI, an assumption necessary to derive the inequalities. We have also described three variants, 
corresponding to different ways in which homogeneity of the initial conditions could be attained: by putting it in by hand ($\text{LPW}_f$), by imposing it via a law-like constraint ($\text{C-LPW}_f$) and by achieving it dynamically ($\text{R-LPW}_f$)---although in this latter case, the model is not truly equivalent to pilot-wave theory. We also introduced a particle version of these models, $\text{LPW}_p$, for which the first two ways of achieving homogeneity seem readily available (regarding the third one, the situation is not clear). We think these models provide a solid platform in which the debate about the feasibility, benefits, and price of rejecting SI can be framed. In the next section, we take the first steps in that process and consider some of the consequences of our models for the general discussion regarding violations of SI.

\section{Discussion}
\label{Dis}
To close up, we start this section by assessing the behavior of our model with respect to the standard worries leveled against superdeterminism. Next, we briefly compare it with 't Hooft's cellular automaton proposal. Finally, we explore a number of lingering, potential objections.

In section \ref{RSI}, we described how certain strong intuitions against non-locality have inspired the superdeterminism program. Likewise, we explored some of the most important intuitions against the viability of such models. We concluded that the whole discussion has been hindered, in part, by the lack of specific models, around which the confrontation of these two sets of intuitions could be examined in detail. Of course, we now have a model that could contribute to the analysis of the issue in significant ways.

Regarding intuitions against superdeterminism, we identified four: 1) that in the absence of workable superdeterministic models, the project loses power; 2) that, even if a model could be constructed, the specification of the required initial conditions would be so complicated that the model would be useless in practice; 3) that superdeterministic models would lack all explanatory power and 4) that they would put in risk all of experimental science. We begin this section by exploring how our model fares regarding these intuitions. 

Regarding the first worry, it is clear that we do have a working local model that is, in fact, able to reproduce all the predictions of standard pilot-wave theory. As for the second, it is also clear that the specification of the initial conditions required to accommodate non-local correlations is not only not that complicated, but that it actually is quite simple. Sure, the initial state required for $\text{LPW}_f$ to reproduce standard pilot-wave theory must be chosen from a measure zero set, but we believe that there is a useful distinction to be made between an initial condition of measure zero and one that is so complicated that we would not even know how to write it down. The latter, but not necessarily the former, would seem to add an additional layer of complexity, making the model useless in practice.

The other two worries are more subtle; let us start with the charge of lack of explanatory power. We saw that the fear is that all explanatory resources of superdeterministic models would boil down to pointing to an extremely unlikely initial state. Moreover, that one completely different such state would be required to explain each separate phenomenon. Is this the fate of our model? It does not seem so. To begin with, it is true, again, that within $\text{LPW}_f$, the necessary initial state must be chosen from a measure zero set. However, any initial state from such a set is able to explain absolutely all violations of the Bell inequality. That is, our model does not require different initial states to accommodate different instances of the violation.

Moreover, we saw that since it is possible to succinctly characterize the set of initial conditions that leads to violations of SI, it is possible to postulate a constraint on the allowable initial conditions. This implies that, for the model with such a constraint, $\text{C-LPW}_f$, the set of initial conditions that leads to violations of SI ceases to be of measure zero---the desired behavior is actually fully generic within the allowed set of initial conditions. This means that the explanation for the breakdown of Bell inequalities within $\text{C-LPW}_f$ does not appear to involve fine-tuning. One might argue that the transition from $\text{LPW}_f$ to $\text{C-LPW}_f$ does not seem like much of an improvement: the distinction between selecting homogeneous initial conditions and enforcing them through a constraint might seem a bit of a cheat. The important point to notice, though, is that this trick is possible only because the required constraint on the initial condition is extremely simple and can be summarized by the succinct expressions in Eqs. (\ref{Hom1}) and (\ref{Hom2}). 

It is also important to mention that the change of measure over the space of initial states, with the introduction of the constraint, is fully analogous to what happens in other theories with constraints, such as gauge theories or general relativity (in the ADM formulation). For instance, in Maxwell's theory, only initial data satisfying $\nabla \cdot E = 4\pi \rho$ and $\nabla \cdot B =0$ are physical, and thus are clearly of zero measure when considered as part of the set of function pairs $ (E(x) , B(x))$. On the other hand, in the case of such theories, the constraints emerge directly from fundamental symmetries. However, in our case, they have been imposed in what seems a completely ad hoc manner, with no independent motivation.

Finally, the model with dynamical relaxation, $\text{R-LPW}_f$, would represent an important improvement because it would predict observable non-local correlations, out of \emph{generic} initial conditions, without the need of introducing a constraint. That is, if it works, the model would locally explain all non-local correlations, completely avoiding the necessity to impose a particular type of initial condition. The problem, as we noted above, is that there are some lingering issues to be clarified: the fact that it involves instantaneous propagation (for which we think a solution ought to exist), the potential of not satisfying quantum equilibrium and the fact that the homogenization at finite times is only approximated, leading to deviations from standard predictions.

The last worry regarding superdeterminism was that such models could jeopardize all experimental science because they would imply the breakdown of assumptions analogous to SI in other experimental scenarios. However, such a fear could be countered if we could explain why, within our model, SI is violated only for particular experimental settings, such as those considered by Bell, and not generically. To explore if our model is able to provide such an explanation, it is instructive to explore the way in which our model violates SI. After all, the model, given homogeneous initial conditions, is empirically equivalent in all respects to standard pilot-wave theory. How is it, then, that one violates SI while the other does not. The answer lies in the fact that, while both models are fully equivalent at the level of the predicted mass density over physical space, they are wildly different at the level of the remaining postulated ontology. This, in turn, leads to a very different characterization of $\lambda$ in each case.

In standard pilot-wave theory, the state of the measured pair is given by an entangled wave function, together with the position of the two particles involved; that is all. As a result, under standard experimental circumstances, one does not expect any correlation between such a state and the settings $a$ and $b$. The situation with our model, on the other hand, is very different. In that case, the theory is fully local and separable. As a result, $\lambda$ is composed of the values of the wave function field $\Phi_\mathbf{x} $ and the positions field $Z_\mathbf{x}$ in the region in which the pair is created. The problem is that such a $\lambda$, even though it is constrained to a small region, clearly is not independent of $a$ and $b$ because, given the homogeneity of the initial conditions, from $\Phi_\mathbf{x} $ and $Z_\mathbf{x}$ on a single point, information on $\Phi_\mathbf{x} $ and $Z_\mathbf{x}$ on all other points, including those related to $a$ and $b$, is available (that is, in our model, $\lambda$, $\lambda_a$, $\lambda_b$ and $\lambda_E$ are not independent; see section \ref{facloc}). Clearly, then, SI is grossly violated.

Let us explore in more detail how Bell inequalities are violated in our model. As we saw, one of the fears associated with violations of SI is its conspiratorial nature and the fact that, given the myriad of convoluted methods employed to select settings, models violating SI would have to involve even more convoluted details to explain the observed correlations. That is, they would have to explain why the state of an entangled pair is correlated with old TV shows or with photons from quasars. Our model has a simple answer for these strange correlations: the state of an entangled pair is correlated with old TV shows or with photons from quasars because, according to our model, everything is correlated with everything! This simple answer to the presence of these mysterious correlations, however, leads to an equally pressing question: if, according to the model, everything is correlated with everything, why don't we notice it? That is, why such non-local correlations are only brought to light through highly sophisticated experimental arrangements?

The answer is similar to the one given by, say, pilot-wave theory for why, even though the theory is non-local, such non-locality does not manifest itself so easily. As we explained above, within the pilot-wave theory, non-locality arises only for groups of particles whose wave function is entangled. Similarly, in our model, even though, at the level of $\Phi_\mathbf{x} $ and $Z_\mathbf{x}$ everything is tightly correlated, these correlations manifest themselves at the mass density level, only when there is entanglement. As we mentioned before, our model is ultra-local. Therefore, the dynamics at each point $\mathbf{x} $ is fully autonomous. Of course, due to the homogeneity of the initial conditions, this fully autonomous dynamics is endlessly reproduced all over space. However, this does not directly translate into observable correlations between different points because the recipe to construct the mass density at each point in space in Eq. (\ref{MD}) implies that different aspects of $\Phi_\mathbf{x} $ and $Z_\mathbf{x}$ determine the value of the mass density at each point $\mathbf{x}$. It is only when entanglement is present that the behavior of mass density at different points ``magically'' coordinates to display correlations at such a level. In particular, in a Bell-type experiment, the entanglement between the pair of particles, and their interaction with the measuring apparatuses, lead to violations of the Bell inequalities. On the other hand, in, say, standard randomized clinical trials, since no entanglement is involved, no miraculous non-local correlations are expected, so it seems that those results can remain solid. We conclude that our model is able to explain why SI is violated in Bell-scenarios, but not generically. 
 
One might think that the observations of the last paragraph open the door to modifications of the model that relax the complete homogenization requirement and restrict that to the realm of those variables that are entangled. However, as one does not know a priori which variables might, through the course of time, get entangled, it seems one has no reliable mechanism to ensure that the model reproduces quantum theory under general conditions, other than to demand full homogenization. A related point has to do with the fact that, generically, one expects things to get more and more entangled with time, so it seems that our previous observation regarding the fact that correlations only show up for entangled systems won't be able to explain the scarcity of such correlations. The point, however, is that, just as in the pool tables example discussed above, one expects such correlations, involving so many degrees of freedom to completely wash out (at least at the practical level) once one focuses on just a few particular variables.
 
Before moving on to explore some potentially problematic aspects of our model, we would like to contrast it with Gerard 't Hooft's cellular automaton program \cite{Hooft2016}. As we explained above, 't Hooft has been trying to come up with a deterministic and local theory that could replace quantum mechanics. What we want to highlight is that our model can be seen as a realization of this project. To see this, imagine the spatial points $\mathbf{x}$ in our model, with their associated degrees of freedom, $\Phi_\mathbf{x}(z_{\mathbf{x}},t)$ and $Z_\mathbf{x}(t)$, to be taken as representatives of the individual cells of the cellular automaton envisioned by 't Hooft. The main differences are i) that our cells are point-like while 't Hooft's cells have some size, and ii) that, while in our proposal each cell is fully autonomous, in 't Hooft's proposal individual cells might interact with neighboring cells. We think these two differences might be erased through a rather simple modification of our model. To deal with the first, simply {\it fatten} the cells by taking them associated not with points $\mathbf{x}$ but with, say, cubes of side $\ell_{\text{P}}$; at the same time, replace in the mass-density described by Eq. (\ref{MD}), the delta function by an integral of such an expression over the corresponding cell. To deal with the second difference, we can have the guiding wave function field $\Phi_\mathbf{x}$ replaced by the corresponding averaged version over neighboring cells.

Finally, note that it is not only that our model realizes 't Hooft's objectives regarding locality, but that two of our paths to achieve homogenization of the initial conditions---the postulation of a law and the dynamical approach---resemble his proposals on how to deny SI, mentioned above. It seems clear, then, that our model can be seen as a full realization of 't Hooft's ideas. Moreover, it appears that any model capable of doing so would have to share with our model the (extreme) feature of having at each cell full information of the whole system (or something very similar). This is so because, as we saw before, one cannot know a priori which variables will get entangled. All this implies that we can use our model to assess the alleged virtues and problems of 't Hooft's ideas. In this vein, we close up by focusing on some objectionable aspects of our model.

To begin with, we note that the ultra-local nature of our model---i.e., the fact that the dynamics of the fundamental degrees of freedom $\Phi_\mathbf{x}(z_{\mathbf{x}},t)$ and $Z_\mathbf{x}(t)$ at each point $\mathbf{x}$ is fully autonomous---clearly implies the presence of an absolute rest frame that allows one to talk about the same point in space at different times (note that an analogous feature arises within 't Hooft's proposal regarding the cells). This unattractive feature might be compared with the claim in \cite{Valentini1997} that a correct reading of standard pilot-wave theory also implies the existence of an absolute rest frame. However, the difference is that, even if Valentini is correct, there is no doubt that Galilean invariance is at least an \emph{effective} symmetry of the standard theory. Our model, in contrast, does not respect Galilean invariance for generic initial conditions. Note however that, since for homogeneous initial conditions, our model is, at the mass density level, equivalent to the standard theory, then our model is consistent with Galilean invariance when homogeneous initial conditions are present. All this is reminiscent of the fact that, even though standard pilot-wave theory requires a preferred foliation of space-time, as long as quantum equilibrium is present, such a foliation is not empirically accessible (note that, in our model, homogenization---plus quantum equilibrium---would also block empirical access to the preferred foliation). 

In sum, our model requires an absolute rest frame, and such a frame is empirically accessible for generic initial conditions. Homogeneous initial conditions, on the other hand, block empirical access to such a frame---as well as, if quantum equilibrium is present, access to the associated preferred foliation. Similarly, standard pilot-wave theory requires a preferred foliation, but quantum equilibrium blocks access to it. Now, in the context of standard pilot-wave theory, it has been proposed that the preferred foliation could be selected dynamically \cite{Durr}. It seems, however, that our model, having the local beables themselves intimately associated with space points, would make it harder for a process of that sort to work. Finally, we note that $\text{LPW}_p$ does not require an absolute space. It is clear, however, that this particle version would not be very useful for the construction of a field theoretical relativistic version (and, as we saw, the dynamical relaxation method for achieving homogeneity does not seem readily available for $\text{LPW}_p$).

The next objection we would like to mention is also related to the ultra-local character of our model. The problem is that the autonomy of the dynamics at each point seems to imply that an epistemic agent composed of a collection of such points would be incapable of acquiring information from the outside, seemingly leading to a profoundly solipsistic scenario. In a world described by our model, Descartes' evil demon would have an easy time deceiving us, with no need to substitute the input acquired by the senses with something else---since there would be no input to begin with. Similarly, a mad scientist would not require to provide electrical impulses to a brain in a vat in order to simulate reality, and it would be easier for a Boltzmann brain to spontaneously form. As a result, our model might seem to be empirically incoherent or self-refuting, with a total disconnect between ontology and epistemology.\footnote{A related worry runs as follows: the fact that in our model each point contains information about the entire universe, together with the fact that the dynamics at each point is ultra-local, invites one to adopt a position in which all there is to the world is a single point. The lack of interaction between the internal systems at different points certainly makes such a posture viable, but that would simply amount to returning to standard pilot-wave, with its standard non-locality. Alternatively, one could take the internal space to be a 3N-dimensional configuration space ${\cal C} $ (as is done in our presentation), in which case the internal dynamics would be fully local (in that space). In that scenario, taking the position that all there is to the world is a single point would lead to David Albert's proposal in \cite{Albert}. However, if one insists on having a theory that is local in ordinary space, then such an option would be unavailable.} 

Note, however, that the situation regarding this last issue is not completely clear. In order to arrive at a conclusion regarding empirical incoherence, a full analysis of the possible existence of epistemic agents within our model, and of their epistemic capacities, would need to be carried out. Needless is to say, that is a formidable task well beyond the scope of this paper. Moreover, we should recognize that epistemic questions of this kind could be raised in the context of many other theories, so we should avoid demanding from our model more than what is usually demanded from others. Still, there are a few things we can say, even in the absence of a full epistemological analysis. If we restrict consideration to the situation where homogeneity prevails, and make use of the fact that, under such conditions, our model coincides with standard pilot-wave theory at the mass density level, we could simply ``pass the buck'' regarding epistemic questions from the former to the latter. That seems like a very promising path, except, perhaps, if we delve more deeply into questions regarding epistemic access to the assumption of homogeneity itself. If, on the other hand, one would want to study the theory in all generality, independently of the homogeneity conditions, then the issue would become extremely murky and the risk of epistemic incoherence would persist.
 
The last potential problem we discuss has to do with what could be seen as an unacceptable, gross enlargement of the ontology. Remember that our model substitutes the standard pilot-wave ontology, consisting of one wave function and one position vector (on configuration space), with a wave function and a position vector for each point $\mathbf{x}$ in physical space (or one for each particle in $\text{LPW}_p$). As we mentioned before, the model could be thought of as containing at each point a supercomputer running a perfect simulation of the entire universe. However, this, by itself, should not be seen as a significant objection. First, proponents of any kind of hidden variable theory are already well disposed towards an expanded ontology. Second, it is true that the magnitude of this particular expansion might seem quite large, but, in fact, similar enlargements of ontology have occurred elsewhere in physics, e.g., in the transition from standard, non-relativistic quantum mechanics to quantum field theory. It is clear, then, that an ontological growth should not be judged alone: it must be weighed against the full background of empirical and theoretical vices and virtues of a theory. That is, it must be weighed, along with many other aspects of a theory, such as conceptual clarity, simplicity, explanatory and unificatory powers, etc. (for instance, it is clear that the ontological expansion in quantum field theory is tied to a spectacular increase of explanatory power).

So, what can be said in this regard about our model? Is the enlargement of the ontology in our case worth its cost in this sort of calculus? We seriously doubt it. The only benefit of our model seems to be that it manages to preserve locality, but it does so by, at the same time, seriously muddling its standing regarding clarity, simplicity, explanatory power, etc. For instance, according to our model, there is a dramatic disparity between the full ontology of the model and the part of the ontology that manifests itself in physical space (via the mass density in Eq. (\ref{MD})). That just seems like an unnecessarily cumbersome way of doing things---particularly in light of other theories that do the same in a much simpler way. It is not clear, then, why anyone would choose our model over, say, pilot-wave theory. It might be argued that, by being local, our model considerably gains in unificatory power because it allows for unification with relativity. This, however, depends on non-locality being incompatible with relativity, an idea which is arguably mistaken (see, e.g., \cite{Maudlin}). Moreover, our model might be local but, as we just saw, it requires an absolute rest frame, so it does not seem to be in good shape regarding its compatibility with relativity (as we saw, $\text{LPW}_p$ does not require an absolute Galilean frame, but does not seem appropriate for a generalization to a field theory). By the way, it is clear that a similar situation holds in connection with 't Hooft's proposal. In that case, the preferred frame would correspond to that defined by the world lines of the individual cells that make up the global cellular automaton. It thus seems that if the aim was to construct a theoretical framework that was fully in compliance with the spirit of relativity, the approach would be self-defeating.

In sum, we have presented a model (and several variations thereof) that are fully local but, by explicitly violating SI via the superdeterministic route, manage to reproduce all the predictions of quantum mechanics (with perhaps the exception of the $\text{R-LPW}_f$ version). Moreover, our model can do so without an unmanageably complicated initial state, without fully losing all explanatory power, and without manifestly rendering all experimental science obsolete. That is, our model seems to evade some of the worst fears previously attributed to a potential construction of an SI violating model via superdeterminism. In particular, within the model it is indeed possible to succinctly characterize the set of initial conditions that lead to violations of SI, reducing the risk of the model losing explanatory power. Moreover, the model explains why SI is violated only in Bell-type scenarios, removing the potential threat to experimental science in general.

Still, our model gives rise to a number of unforeseen difficulties, like the risks associated with a possible disconnect between ontology and epistemology or the enormous ontological increase, without any gain in explanatory power. We conclude that the model simply ``feels unnatural'' and does not seem to possess the theoretical virtues that would render it a serious contender to its non-local counterparts. In our view, the price to pay to bypass the non-locality conclusion via this type of rejection of SI simply seems too high. Independently of this conclusion, and having argued that local models that reproduce all predictions of quantum theory via the superdeterministic route must share many of the features of our model, we are convinced that the latter can be very useful in providing a framework for further discussion of all these issues.

\section*{Acknowledgments}
We would like to thank Tim Maudlin for extensive and illuminating conversations on the subject of this work. We also thank Travis Norsen and Ward Struyve for valuable comments and two anonymous referees, whose criticisms and comments contributed to improving this paper over its first version. D.S. acknowledges partial financial support from PAPIIT-UNAM grant IG100120; the Foundational Questions Institute (Grant No. FQXi-MGB-1928); the Fetzer Franklin Fund, a donor advised by the Silicon Valley Community Foundation. E.O. acknowledges support from UNAM-PAPIIT grant IN102219. G.S.C acknowledges support from a CONACYT scholarship.
\bibliographystyle{plain}
\bibliography{bibbell.bib}
\end{document}